\documentclass[12pt,preprint]{aastex}
\newcommand{\au}{\rm\ AU}

\newcommand{\kelvin}{\rm\ K}
\newcommand{\years}{\rm\ yr}
\newcommand{\percubiccm}{\rm\ cm^{-3}}
\newcommand{\ergpcc}{\rm\ erg\percubiccm}
\newcommand{\solsperyear}{\rm\ M_\odot\ yr^{-1}}
\newcommand{\kmpersec}{\rm\ km\ s^{-1}}
\newcommand{\rdisk}{r_{\rm d}}
\newcommand{\rstar}{r_{\rm s}}
\newcommand {\bmax}{B_{\rm m}}
\newcommand {\rmax}{r_{\rm m}}
\newcommand{\betamax}{\beta_{\rm m}}

\newcommand{\nambient}{n_{\rm a}}
\newcommand{\tambient}{T_{\rm a}}
\newcommand{\tfloor}{T_{\rm f}}
\newcommand{\pwind}{P_{\rm w}}
\newcommand{\rhowind}{\rho_{\rm w}}
\newcommand{\vwind}{v_{\rm w}}
\newcommand{\rwind}{r_{\rm w}}
\newcommand{\vstellar}{v_{\rm s}}
\newcommand{\vdisk}{v_{\rm d}}
\newcommand{\mdotdisk}{\dot M_{\rm d}}
\newcommand{\mdotstellar}{\dot M_{\rm s}}
\newcommand{\thetadisk}{\theta_{\rm d}}
\newcommand{\thetastellar}{\theta_{\rm s}}
\newcommand{\thetawind}{\theta_{\rm w}}
\newcommand{\magpower}{\mathcal{P}_{\rm B}}
\newcommand{\totalpower}{\mathcal{P}_{\rm tot}}
\shorttitle{Magnetized, nested PPN winds}
\begin{document}
\title{Magnetic Nested-wind Scenarios for Bipolar Outflows: Pre-planetary and YSO nebular shaping
\footnote{The data presented in this paper were obtained from the Multimission 
Archive at the Space Telescope Science Institute (MAST). STScI is operated by the 
Association of Universities for Research in Astronomy, Inc., under NASA contract 
NAS5-26555. Support for MAST for non-HST data is provided by the NASA Office of 
Space Science via grant NAG5-7584 and by other grants and contracts.}}
\author{Timothy J. Dennis\altaffilmark{1}, Adam Frank\altaffilmark{1}, Eric G. Blackman\altaffilmark{1}}
\author{Orsola DeMarco\altaffilmark{2,5}, Bruce Balick\altaffilmark{3}}
\author{\and Sorin Mitran\altaffilmark{4}}
\altaffiltext{1}{Department of Physics \& Astronomy, University of Rochester, Rochester, NY 14627; tdennis@pas.rochester.edu}
\altaffiltext{2}{ Department of Astrophysics, American Museum of Natural History, Central Park West at 79th Street, New York, NY 10024}
\altaffiltext{3}{Department of Astronomy, University of Washington, Seattle, WA 98195}
\altaffiltext{4}{ Department of Mathematics, University of North Carolina, Chapel Hill, NC 27599}
\altaffiltext{5}{Department of Physics, Macquaurie University, Sydney NSW 2109, Australia}
\begin{abstract}
We present results of a series of magnetohydrodynamic (MHD) and hydrodynamic 
(HD) 2.5D simulations  of the morphology of outflows driven by nested wide-angle winds - i.e. winds which eminate from a central star as well as from an orbiting accretion disk. While our results are broadly relevent to nested wind systems we have tuned the parameters of the simulations to touch on issues in both Young Stellar Objects and Planetary Nebula studies.  In particular our studies connect to open issues in the early evolution of Planetary Nebulae.
We find that nested MHD winds exhibit marked morphological differences from the single MHD wind case along both  dimensions of the flow. Nested HD winds on the other hand give rise mainly to geometric 
distortions of an outflow that is topologically similar to the flow arising from a 
single stellar HD wind. Our MHD results are insensitive to changes in ambient temperature 
between ionized and un-ionized circumstellar environments. The results are sensitive to the 
relative mass-loss rates, and to the relative speeds of the stellar and disk winds. We also present synthetic emission maps of both nested MHD and HD simulations. We find that nested MHD winds show
knots of emission appearing on-axis that do not appear in the HD case.
\end{abstract}
\section{Introduction}
\label{sec:intro}
Starting in 1994, the Hubble Space Telescope (HST) began returning images
with unprecedented detail of what were thought to be familiar and well-understood 
objects. The images made it clear that the so-called Generalized Interacting 
Stellar Winds (GISW) model was inadequate to explain the morphologies of several 
classes of Planetary Nebulae (PN's) and Proto-Planetary Nebulae (PPN's) 
(Balick \& Frank 2002). In particular the narrow-wasted bipolar, multipolar, 
and point-symmetric classes could not be explained as being solely due to the interaction
of an isotropic fast wind with the material previously deposited by the slow, ``superwind'' 
phase of AGB mass-loss; even when accounting for the ubiquitous presence of dense,
dusty disks or tori surrounding the central post-AGB star first observed by Balick (1987). 
In addition, a number of smaller-scale, low-ionization, observational features 
associated with approximately half of known PN's in the form of knots, collimated 
jet-like structures, and the remarkable ``Fast Low-Ionization Emission Regions'' 
(FLIERS) indicate that there is much that remains to be understood about 
the physics occurring in these environments (Gon\c{c}alves et al. 2001).

An H$\alpha$ HST survey of very young PN's carried out by Sahai \& Trauger (1998)
revealed the presence of bipolar ansae and/or collimated radial structures indicating
the presence of jets. In others, bright structures in proximity to the minor axes were 
observed, indicative of the disks and tori mentioned previously. Sahai \& Trauger (1998) 
proposed that high-speed collimated jets serve as the primary agent of the shaping 
process. Later theoretical work by Soker \& Rappaport 2000 and Soker 2004 entensively explored the ability of collimated hydrodynamic winds to shape PN.  Numerical work on this model has been carried out in a series of revealing studies by Lee \& Sahai (2003, 2004, 2009) and Akashi \& Soker (2008)

The presence of both disks and jet-like structures is reminiscent of the enviroments 
of Young Stellar Objects (YSO's), where jets drive ubiquitous molecular outflows.  YSO 
jets are believed to be magnetically launched and these mechanisms for driving the outflows have been extensively studied (Blandford \& Payne 1982, Pudritz \& Norman 1983, 
Uchida \& Shibata 1985, Shu et al. 1988, Contopoulis \& Lovelace 1994, Shu et al. 1994a,b, 
Goodson \& Winglee 1999).   This led some to speculate that jets in PN's and 
PPN's may also be magnetically driven (Blackman et al. 2001a, 2001b; Frank \& Blackman 2004; Matt et al. 2006; 
Frank 2006). Some indirect support for MHD driving in PN's is provided by the observations of 
Bujarrabal et al. (2001) who, in a survey of 37 PPN's, found fast winds associated with 28
of these objects with momenta that are in most cases too high---sometimes by a factor
of $~10^3$---to be accounted for by radiation pressure.  

Blackman et al. (2001a) have shown that a dynamo operating in an AGB star can produce magnetic 
fields powerful enough to drive a self-collimating outflow accounting at once for both
the momentum problem and the observed collimation. The means by which the field
so generated drives and collimates the outflow is presumed to be the ``magnetocentrifugal
launch'' mechanism (MCL) (Blandford \& Payne 1982, Pelletier \& Pudritz 1992). 
Because such a dynamo operating in an isolated star is subject to the criticism that 
some mechanism for restoring shear is necessary in order to maintain it, more recent
work has focused on common-envelope dynamos in which the rotational energy needed to maintain
the dynamo is supplied by an embedded low-mass companion. Nordhaus et al. (2007) have 
recently shown that for a variety of such scenarios a robust dynamo results.

In the MCL scenario for disk-winds, a poloidal field threading the disk and 
sufficiently inclined with respect to the disk axis, acts as a conduit for coronal gas 
experiencing centrifugal forces that overbalance gravitational attraction. The 
material is thus ``flung'' out along the poloidal lines of force, until it passes beyond 
the ``Alfv\`en surface'', where magnetic tension is no longer sufficient to maintain 
co-rotation of the field resulting in shear and the consequent development of a toroidal 
component which ultimately dominates at sufficiently large distances.  The toroidal field 
is buoyant and thus ''rises'' through the radially stratified circumstellar environment 
carrying the disk material with it while simultaneously ensuring a high degree of 
collimation of the material it carries because of the field's hoop stress. 

It has additionally been pointed out that the MCL scenario can occur more impulsively
even in the absence of disks when linked to the rapid evolution of the source of
the field. The driver in this case would be stellar differential rotation and the consequent 
shearing of a stellar magnetic field in the ionized circumstellar environment. 
Such a mechanism may apply to gamma-ray bursts or supernovae (e.g. Piran 2005), and has 
been examined both analytically and numerically by a number of authors 
(Klu\`zniak \& Ruderman 1998; Wheeler et al. 2002; Akiyama et al. 2003; Blackman et al. 2006). 
The added presence of disks in the environments of post-AGB stars, together with the presumption 
of the presence of magnetic fields in both, have led us to consider the possibility that not 
one, but two ``fast'' ($10^2-10^3 \kmpersec$) magnetic winds may be operating simultaneously 
in the enviroments of at least some post-AGB objects. In such a scenario it is likely that the 
spherical symmetry of the stellar-wind source will result in a broader opening angle than that 
of the disk-wind that surrounds it, leading to a collision of the two winds occurring relatively 
near their sources. Such an interaction, if it occurs, is very likely to have a profound influence 
on the ensuing morphological development of the shock-heated emitting structures thereby manifested. 

Ro\`zyczka \& Franco (1996) were the first to present simulations
showing that a diverging fast-wind, threaded by a toroidal magnetic field, and 
incident upon an unmagnetized ambient medium modeled to be consistent 
with environments observed around evolved stars, can collimate the wind provided
the magnetic field is sufficiently strong. Soon after, Frank et al. (1998) presented a set 
of axially symmetric magnetohydrodynamic (MHD) simulations  examining the influence of 
strong magnetism on the morphological and kinematical features of radiative jets in the 
context of YSO's, and found that they differ significantly from corresponding hydrodynamic (HD) and 
weak field cases, forming ``nose-cones'' at the head of the jet, narrower bow shocks, 
and enhanced bow shock speeds. These effects were attributed to the hoop stresses 
imposed upon the flow by the toroidal field. In subsequent work, Frank et al. (2000)
added greater realism by using analytical models of MCL 
launching to specify the cross-sectional distributions of the jet's state variables.
The resulting radial stratification of density and magnetic field led to new propagation
behavior manifested principally by the development of an inner jet core within a 
lower density collar. Several studies of both pulsed and steady radiative jets have
addressed the effect of various magnetic field topologies on the emission 
features of jets (Stone \& Hardee, 2000; O'Sullivan \& Ray, 2000; 
Cerqueira \& de Gouveia Dal Pino 1999, 2001a,b; de Gouveia Dal Pino \& Cerqueira 2002). 
It is found that the emission structures resulting from the imposition of helical 
or toroidal field configurations depart the most from their purely hydrodynamic 
analogs, but that the differences that arise become less pronounced, and in the case 
of the nose cones, even vanish in fully three dimensional calculations
(Cerqueira \& de Gouveia Dal Pino 1999, 2001a,b). De Colle \& Raga (2006) 
have studied the H$\alpha$ emission of  axisymmetric radiative jets threaded 
with toroidal fields and have concluded that the greater jet collimation leads to
an increase in H$\alpha$ emission along the jet axis, and, like Frank et al. (1998), a
somewhat increased shock velocity relative to the hydrodynamic case.

The possibility of contemporaneous disk and stellar winds is not restricted
to PN's and PPN's. Evidence exists for this phenomenon in every environment 
where jets and accretion disks are found (see, e.g. Livio 1997 and references therein). 
Nonetheless most studies of magnetized winds focus on individual jets and do not 
address the question of how simultaneously operating stellar and disk winds may 
interact with one another as evidenced by the brief survey of literature provided above. 
The relatively few studies involving simultaneous outflows include those of  
Meliani et al. (2006) and Casse et al. (2007), who examined the launch physics of 
simultaneous outflows in the YSO context using non-ideal MHD simulations for a 
self-consistent accounting of the viscous and resistive accretion disk, of 
Matsakos et al. (2008), who studied the topological stability of two-component outflows 
for a pair of prototypical and complementary analytical solutions via time-dependent 
MHD simulations, and of Fendt (2008), who addressed the question of how the formation of
large-scale jets is affected by the interaction of the central stellar magnetosphere and 
stellar wind with a surrounding magnetized disk outflow using axisymmetric MHD simulations. 

In the context of PN's, the ability of collimated winds to produce the diverse features seen in many collimated PN's has been studied in some detail in the hydrodynamic case by Akashi (2007), Akashi \& Soker (2008) and Akashi et al (2008). In these studies a wide angle wind ($\theta> 10^o$) from a central source was ejected into a spherical AGB wind.  The evolution of the subsequent nebula was tracked to observable scales.  The authors showed that the resulting morphology could recover a number of important features seen in real PN's such as front lobes and rings on the main bipolar structure.  Of particular importance were the equatorial rings formed as wind-angle winds/jets would lead to compression of material in the symmetry plane (which is expected to be both the plane of the disk and the plane of the binary orbit.  These studies were important in their ability to demonstrate the range of features which could be produced via wide angle jets.

We are presently unaware of any previous numerical study examining the 
macroscopic features of the resulting flow and environment for two simultaneously flowing, 
nested winds.  The purpose of the present work is to examine the morphological 
consequences of pairs of simultaneous, steady, radiative, 
toroidally magnetized, and nested winds, and to compare the results of these simulations 
to similar winds for which the magnetic field and/or the disk wind is absent. 

The remainder of this paper is structured as follows: section \ref{sec:numericalmodel}
presents a description of the simulations performed, detailing the boundary
conditions used for the wind-launching region, the parameters chosen to characterize
the winds, the ambient environment into which they flow, the nature of the magnetic 
field imposed and the initial conditions for each simulation presented.
In section \ref{sec:res} we present graphical comparisons among the various 
simulations and our description and interpretation of the structures observed. 
In section \ref{sec:conc} we present our conclusions.

\section{Numerical Model}
\label{sec:numericalmodel}
\subsection{Geometry}
\label{sec:geometry}
\begin{figure*}[ht]
\includegraphics[angle=0,
		 width=5.4in,
                 keepaspectratio=true,
                 trim= 0 0 0 0
                 clip=true]
        {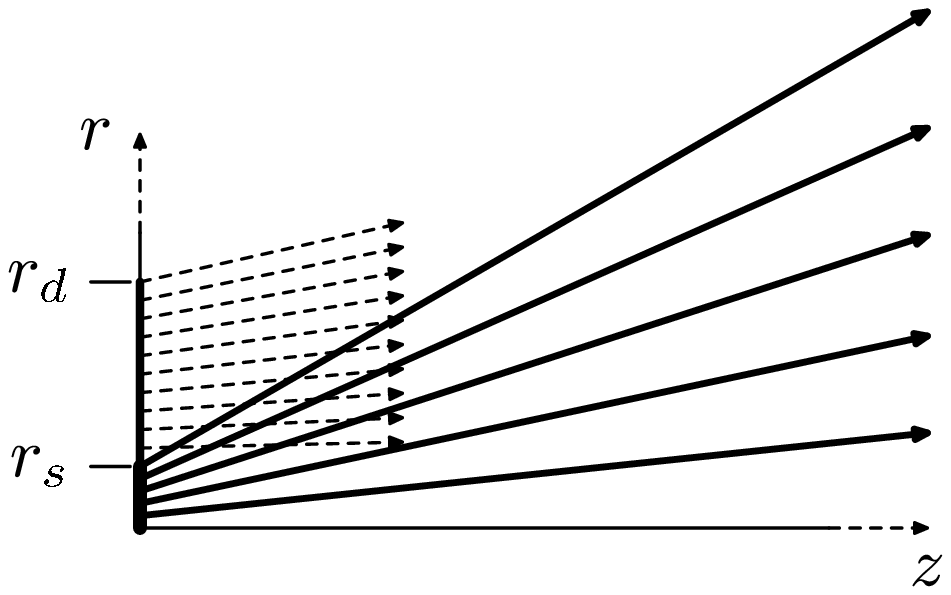}
\caption{Boundary condition for the lower-left corner of the 
computational domain. The solid arrows represent velocity vectors for the inner wind, while
the dashed arrows represent velocity vectors for the disk wind. The outer-most solid arrow
makes an angle of $30^\circ$ with respect to the horizontal axis, while the outer-most dashed
arrow makes an angle of $15^\circ$ with respect to the vertical axis. These values represent the
stellar wind opening angle and the disk wind opening angle respectively and are used in all
simulations presented in this paper. Also for all simulations we set $\rdisk=500\au$ and 
$\rstar=125\au$ respectively.  (See section \ref{sec:numericalmodel} for a detailed description
of the launch region boundary condition.) 
\label{fig:f1}}
\end{figure*}

We execute a series of axisymmetric, radiatively cooled
simulations of two co-axial, or ``nested,'' 
magnetohydrodynamic (MHD), and hydrodynamic (HD) winds 
flowing simultaneously into a rectangular domain from a source 
located at the lower left boundary. The long ($z$) axis is 
chosen to correspond to a physical size equal to $2\times10^4 \au$ 
while the short ($r$) axis is $8\times10^3\au$. The 
direction of flow is predominantly along $z$. A diagram 
of the boundary condition assumed in the region where the
winds are launched is shown in figure \ref{fig:f1}. Two 
radii, $\rdisk$ and $\rstar$, are defined to delineate 
the regions where the inner ``stellar'' wind and the 
outer ``disk'' wind operate. In the range $ r<\rstar$ 
of the launch region, the parameters specifying the 
stellar wind determine the properties of the flow there, 
while for $\rstar<r<\rdisk$ the flow properties are 
determined by the parameters specifying the 
disk wind. The parameters chosen to specify the winds 
include the disk-wind and stellar-wind mass-loss rates; 
$\mdotdisk$, and $\mdotstellar$; their velocities; 
$\vdisk$ and $\vstellar$; their opening angles; 
$\thetadisk$ and $\thetastellar$, and two dimensionless 
parameters $\betamax$, and $\sigma$ to be discussed below. 
As indicated in figure \ref{fig:f1} the wind opening 
angles are upper limits on the degree to which the flow
directions depart from being parallel with the $z$-axis.
For a cell in the launch region a distance $r$ from
the origin, the velocity vector for the flow emerging from
that element makes an angle $(r/\rwind)\thetawind$ with 
respect to the $z$-axis, where $\thetawind$ is one of $\thetastellar$
or $\thetadisk$, and where $\rwind$ is one of $\rstar$ or $\rdisk$ 
depending on whether $r<\rstar$ or $r>\rstar$
respectively.  The opening angles are both non-zero with 
the disk-wind opening angle shallower than that of the stellar 
wind.  The intent is to choose opening angles here that accord 
with the notion that the disk wind is launched 
magnetocentrifically and ``flung'' out along poloidal 
field lines while the stellar wind mass loss is more 
nearly isotropic.  The angles chosen for the simulations 
presented here represent our best effort to model this 
scenario while respecting the technical limitations 
imposed upon us by the code.
\subsection{Parameterization}
\label{sec:parameterization}
Throughout the launch region, at each time-step, a toroidal 
magnetic field is embedded in the winds. To characterize 
the strength and dynamical significance of this field we introduce the 
independent parameters $\sigma$ and $\betamax$, where 
$\sigma$ is the ratio of wind magnetic energy density
to wind kinetic energy or,
\begin{equation}
\label{eq:sigmadef}
\sigma=\frac{B^2/8\pi}{\rhowind\vwind^2},
\end{equation}
and $\betamax$ is a particular value of the ratio of thermal 
pressure to magnetic pressure, chosen to be characteristic of the wind, i.e.,
\begin{equation}
\betamax=8\pi\pwind/\bmax^2,
\end{equation}
where $\bmax$ is a maximum value for the magnetic field strength.
By specifying the values of $\sigma$ and $\beta$, and 
constraining the value of wind mass density,
$\rhowind$, by requiring a relation of the form
\begin{equation}
\label{eq:MdotConst}
\dot M = \Omega\rhowind \vwind \rwind^2
\end{equation}
(where $\Omega$ is the solid angle of the wind) to hold among the mass-loss rates, velocities, and radii 
of the stellar and disk winds \footnote{for the disk wind 
we require $\rwind^2=\rdisk^2-\rstar^2$}, we infer a
characteristic value for the thermal pressure of the 
wind $\pwind$ by way of the relation
\begin{equation}
\pwind=\rho_{w}\vwind^2\sigma\beta,
\end{equation}
and fix the value of $\bmax$, from the definition of $\betamax$. I.e.,
\begin{equation}
\bmax=\sqrt{8\pi\pwind/\betamax}.
\end{equation}
Finally, using the values of $\pwind$ and $\bmax$ thus obtained
we model the magnetic and pressure profiles, $B(r)$ and $P(r)$
for the winds after the form first introduced by 
Lind~et~al.~(1989). For the magnetic field profile we write:
\begin{equation}
\label{eq:fieldprof}
B\left(r\right)=\Biggl\{
\begin{array}{cc}
\bmax\frac{r}{\rmax}, & 0\le r<\rmax \cr
\bmax\frac{\rmax}{r}, & \rmax\le r<\rdisk \cr
0, & r\ge\rdisk
\end{array},
\end{equation}
and for the pressure profile we write:
\begin{equation}
\label{eq:presprof}
P\left(r\right)=\Biggl\{
\begin{array}{cc}
\left[\alpha+\frac{2}{\betamax}\left(1-\frac{r^2}{\rmax^2}\right) \right]\pwind, & 0\le r<\rmax \cr
\alpha\pwind, & \rmax\le r<\rdisk \cr
\pwind, & r \ge \rdisk
\end{array}.
\end{equation}
The quantity $\alpha$ is a constant related to $\betamax$ according to:
\begin{equation}
\label{eq:defalpha}
\betamax^{-1}=\left(1-\alpha\right)\left(\rdisk/\rmax\right)^2,
\end{equation}
and $\rmax$ is the value of $r$ at which $B=\bmax$, and is chosen 
in all of the simulations presented here to be equal to the 
``stellar'' radius  $\rstar$. One may verify that with the profiles
defined as in equations (\ref{eq:fieldprof}) and (\ref{eq:presprof})
the pressure and magnetic field satisfy the condition of 
magnetostatic equilibrium, i.e., 
\begin{equation}
\label{eq:stateq}
\frac{dP}{dr}=-\frac{B}{4\pi r}\frac{d(rB)}{dr},
\end{equation}
everywhere in the interior of the winds except at $r=\rmax$ where the value 
of $\pwind$ (and therefore $\bmax$) changes. The winds are launched into a 
homogeneous, unmagnetized ambient medium whose total number density, $\nambient$, 
and temperature $\tambient$ are chosen independently. As a consequence, the 
winds are not pressure-matched with respect to their environment in the 
simulations. We allow this freedom so that we may explore how shaping is 
affected by changes in the circum-stellar environment (see section \ref{sec:cooljets}).

Lastly we note that since all of the parameters $\sigma$, $\betamax$, 
$\mdotstellar$, $\mdotdisk$, $\vdisk$, $\vstellar$, $\rdisk$, and $\rstar$
are set independently, the simulations are not controlled with respect
to the total power in the winds. However, because the field strength
in the winds has been kept relatively low, the magnetic contribution
to the outflow power, $\magpower$, is small 
($\magpower\lesssim\sigma\totalpower$)  in all of the 
simulations presented, facilitating the comparisons below and 
arguing against the possibility that the differences seen between the 
MHD and HD runs are the result of a dominant energy effect.

\subsection{Methods and initial conditions}

We have executed nine, radiatively cooled, axially-symmetric simulations 
using ``AstroBEAR.'' AstroBEAR is an AMR Hydro/MHD code based on the conservative form 
of the MHD equations and designed for use with high-resolution shock capturing 
methods for sets of nonlinear hyperbolic equations.  We use AstroBEAR to solve 
either the ideal MHD equations for the magnetized winds, or the equations 
of inviscid hydrodynamics (i.e. the Euler equations) for the unmagnetized 
winds. In both cases we assume that the field variables do not depend on 
the azimuthal angle so that our solutions are axially-symmetric. This 
allows us to reduce the problem to a 2-dimensional calculation.  Because the 
solutions that result represent ``slices'' through the axis of symmetry of 
the full 3-dimensional axisymmetric solution, they are said to be 
2.5-dimensional (2.5D).  For a description of AstroBEAR and the equation set 
it solves see Cunningham, et al. (2008). 

Short descriptions of each simulation and their 
parameterizations are given in table's \ref{tab:t1} and 
\ref{tab:t2}.  Table \ref{tab:t1} lists the values of those
parameters which are common to all cases, and table
\ref{tab:t2} lists each run and its corresponding 
parameterization. All but one of these simulations was 
carried out on a grid of base resolution $640\times256$ and 
with two levels of refinement for an effective resolution 
$2560\times1024$.  One simulation (run A2) was carried out 
on a fixed grid of resolution $2560\times1024$. In every 
case the discretization used corresponds to a resolution of 
64 cells per disk-wind radius, or 16 cells per stellar-wind 
radius. The standard case, in which two MHD
winds interact as suggested by figure \ref{fig:f1}, is 
presented as run A1. For purposes of comparison, we also 
present a hydrodynamic case, cases for which the outermost
of the two winds is absent, and cases for which the 
stellar wind is ``lighter'' than the disk wind, (i.e., the 
stellar mass-loss rate is an order of magnitude smaller).
Two further cases are also presented for nested HD and MHD
winds in order to compare the morphologies obtained in 
a warm ($\sim 10^4 \kelvin$) circumstellar environment
to those obtained in a cool ($\sim 10^2 \kelvin$) environment.

The domain of computation is scaled so that one 
computational unit of length corresponds to $500 \au$. The 
physical size of the domain thus corresponds to a length 
along $z$ of $2\times10^4\au$ and a length along $r$ of
$8\times10^3\au$. The simulations are each run for the 
length of time required for the flow to reach the right
end of the domain. These times are given in physical units
in table \ref{tab:t2}.

We note that the use of a constant density ambient medium was a response to our desire to make the simulations as general as possible allowing these models to address issues relevent to both YSO jets, PNe and other disk/central source systems.  Not including a $1/r^2$ density fall off appropriate to the AGB wind will affect the results in terms of timescales (as the winds/jets will see lower momentum densities in the ambient medium as they propogate outwards).  If the AGB wind is spherically symmetric on nebular scales we do not, however, expect dramatic changes in morphology. This point can be explored in further studies which are beyond the scope of the current work.  We note some aspects of this problem have been covered by Akashi 2007, and Akashi \& Soker 2008 for the nonmagnetic case.

We note also that the velocity scales are chosen to be appropriate for either YSOs (a wind from the inner edge of a disk) or the preplanetary nebular phase. For the preplanetary case we take this speed to be indicative of the bridge between the AGB wind (10 km/s) and the circumstellar wind (1000 km/s). In addition we note that we are interested in the long term of evolution of the morphology and so we have had to make certain choices based on computational expediency and our desire to provide simulations that are generally relevent to nested wind systems. Thus the size scale of the disk boundary condition is larger than should be expected in some systems.  Future studies will be needed to connect behavior at the smallest scales where the disk winds are launched and disk and stellar winds interact (Garcia- Arrendondo \& Frank 2006) and the largest scales where full nebular morphology has been been established.  

While our initial ambient temperature is appropriate for mature PNe (see Akashi \& Soker 2008) it is too high for PPN and YSOs except irradiated YSO jets where $T = 10^4 K$ is a reasonable choice for the ambient medium.  As we shall demonstrate however in the case of MHD winds, which are our principle concern, the choice of the ambient temperature is not a significant factor in the determining the morphology. We have included a discussion early in the paper on the distiction between YSO and PN temperatures.

We use relatively wide jets ($\theta>10^\circ$) in our simulations. The presence of such a wide jets or wide-angle winds has been conjectured for some in time in YSO systems (Shu et al 1994) and can be seen as a diagnostic for launch mechanisms.  In PN systems Soker 2004 has presented analytical arguments for the existence of such wide winds/jets.  Such wide outflow systems are likely to be important for creating wider lobes in the observed nebulae in both classes of bipolar nebulae (YSO and PN) .

In all cases, optically-thin, atomic line cooling based
upon the cooling curve of Dalgarno and McCray (1972) is
assumed. A temperature ``floor,''
$\tfloor=9.0\times10^3$, is set so that only material 
heated to temperatures $T\ge\tfloor$ is subject to
radiative energy loss. No attempt is made to follow 
the ionization dynamics of the flow. Note that adiabatic cooling continues to operate below the floor for radiative cooling. Because our focus 
in this paper is restricted to the morphological features 
that arise from the interaction of nested winds, and from 
their interaction with their common environment, this 
approximation is not expected to materially affect our 
conclusions.

\begin{table*}[ht]
 \scriptsize
\begin{tabular}{ll}
\hline
\hline
Effective Resolution\dotfill & 16 cells/$\rstar$ \\
$\sigma$\dotfill & 0.1 \\
$\beta$\dotfill & 1.0 \\
$\nambient$\dotfill & $5\times10^3\percubiccm$ \\
$\thetastellar$\dotfill & $30^\circ$ \\
$\thetadisk$\dotfill & $15^\circ$ \\
$\rstar$\dotfill & $125\au$ \\
$\rdisk$\dotfill & $500\au$ \\
\hline
\end{tabular}
\caption{\normalsize Parameters common to all simulations
\label{tab:t1} }
\normalsize
\end{table*}

\begin{table*}[ht]
 \scriptsize
\begin{tabular}{llllllll}
\hline
\hline
\vspace{0.1cm}
\\
 Run & Description & $\mdotstellar$ & $\mdotdisk$    & $\vstellar$ & $\vdisk$    & $\tambient$ & run time              \\
     &             &($\solsperyear$) & ($\solsperyear$) & ($\kmpersec$) & ($\kmpersec$) & ($\kelvin$)   & ($10^3\years$)  \\
\hline
\\
 A1   & MHD, both winds on                       & $10^{-7}$  & $10^{-7}$    
     & $150$                          & $50$    & $10^4$     & $1.3$     \\
 B1   & HD, both winds on                     & $10^{-7}$  & $10^{-7}$       
     & $150$                          & $50$    & $10^4$     & $3.1$     \\
 A2   & MHD, disk wind off                      & $10^{-7}$  & $0$           
     & $150$                          & $0$   & $10^4$     & $0.9$     \\
 B2   & HD, disk wind off                    & $10^{-7}$  & $0$              
     & $150$                          & $0$     & $10^4$     & $3.0$     \\
 A3   & cool MHD, both winds on                  & $10^{-7}$  & $10^{-7}$    
     & $150$                          & $50$    & $10^2$     & $1.3$     \\    
 B3   & cool HD, both winds on                & $10^{-7}$  & $10^{-7}$       
     & $150$                          & $50$    & $10^2$     & $3.9$     \\   
 A4   & MHD, light stellar wind                    & $10^{-8}$  & $10^{-7}$  
     & $150$                          & $50$    & $10^4$     & $2.6$     \\
 A5   & MHD, light stellar wind, eq. vel.'s      & $10^{-8}$  & $10^{-7}$    
     & $100$                          & $100$   & $10^4$     & $1.7$     \\
 A6   & MHD, light stellar wind, eq. vel.'s      & $10^{-8}$  & $10^{-7}$    
     & $100$                          & $100$   & $10^2$     & $1.6$     \\
\hline                                                                                             
\\
\vspace{-0.35cm}
\\
\hline
\hline
\end{tabular}
\caption{\normalsize Simulation 
                     Parameters
\label{tab:t2} }
\normalsize
\end{table*}

\section{Simulation results}
\label{sec:res}
\subsection{Density comparisons}

\begin{figure*}[ht]
\includegraphics[angle=0,
		 width=4.8in,
                  keepaspectratio=true,
                  trim= 0 0 0 0
                  clip=true]
         {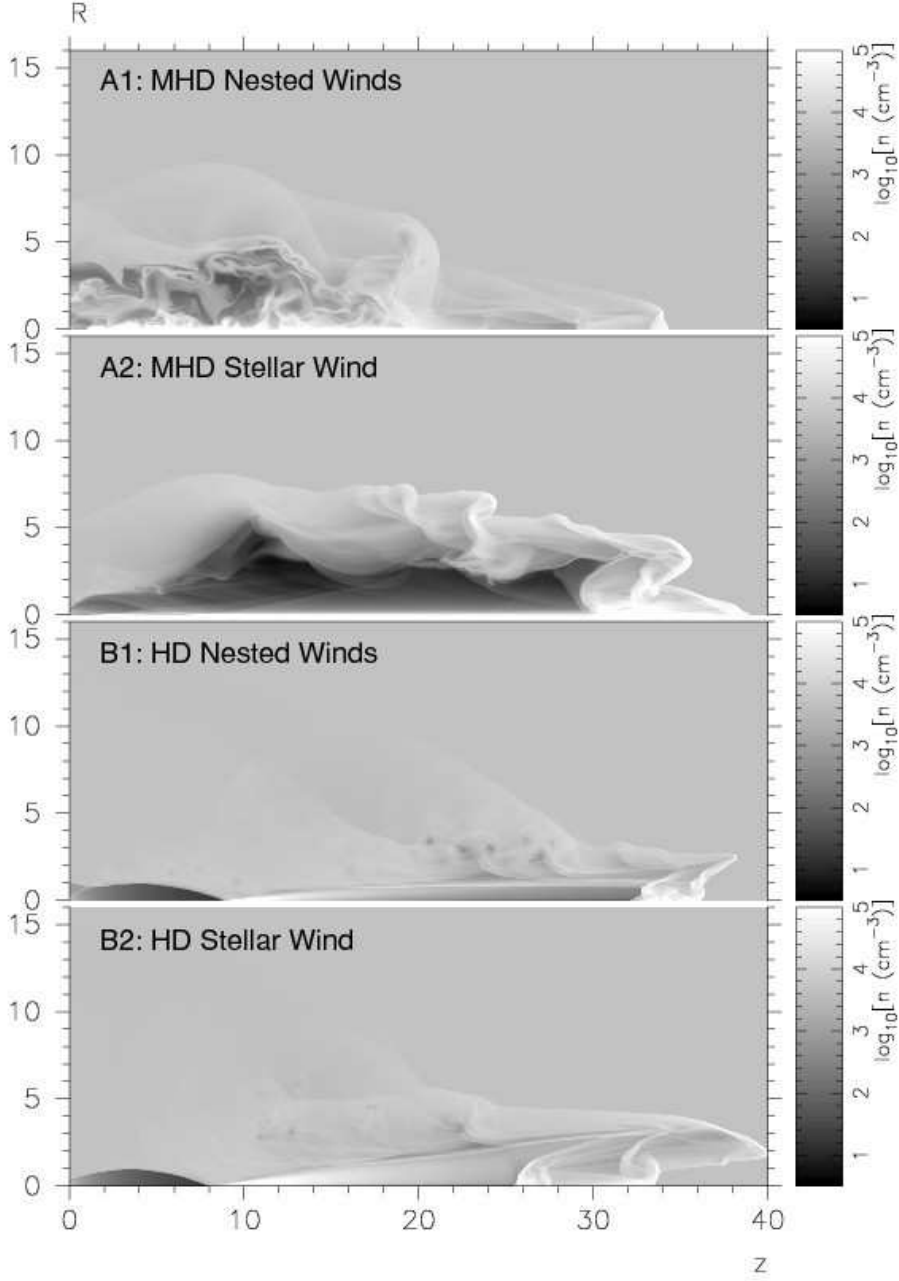}
 \caption{Density grey-scales comparing nested-wind simulations to simulations with only a stellar 
          wind for the MHD and HD cases. From top to bottom, the first 
          and second panels show the MHD case with both winds operating and 
          with only the stellar wind operating respectively (runs A1 and A2 in 
          table \ref{tab:t1}). Similarly, the third and fourth panels show the HD 
          simulations for the case of both winds operating, and with only the stellar 
          wind operating respectively (runs B1 and B2). In all panels, the unit of length is $500 \au$ 
          and the flow axis is parallel to the image plane.
 \label{fig:f2}}
 \end{figure*}

\begin{figure*}[ht]
 \includegraphics[angle=0,
 		 width=5.0in,
                  keepaspectratio=true,
                  trim= 0 0 0 0
                  clip=true]
         {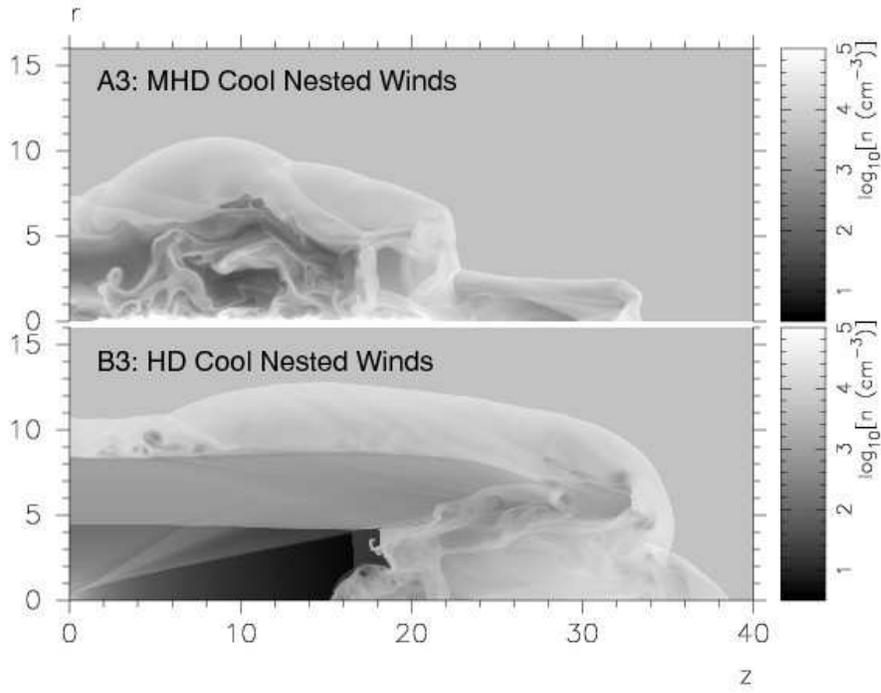}
 \caption{Comparison of MHD (top panel) and HD (bottom panel) nested 
          wind flows for the case of a cool,($T_a = 100 K$), ambient medium. Both stellar
          and disk winds operate for both of the cases shown, (runs A3 and B3 respectively).
          In both panels the unit of length is $500\au$ and the flow axis is parallel to 
          the image plane.
 \label{fig:f6}}
 \end{figure*}

We present the results of our simulations in figures \ref{fig:f2}$-$\ref{fig:f8}.
In figure \ref{fig:f2} we present late-time density maps for runs A1, A2, B1 and B2. 
\footnote{The specific run times to which all maps shown in this paper correspond are given for each
simulation in table \ref{tab:t2}.} The intent here is to contrast the case of simultaneous 
nested winds, with the case of the stellar wind in the absence of the disk wind, for both the MHD, 
(runs A1 and A2) and HD (runs B1 and B2) cases.  Collimation is evident in all 
four simulations. For the HD cases this is a consequence of the ram pressure of the 
ambient medium. For the MHD flows, the hoop stresses associated with the toroidal field 
will also contribute. 

The presence of the slower disk wind does not appear to have as 
significant an effect on the resulting structure of the flow in the case of the hydrodynamic 
winds. In both cases the stellar wind is refocused toward the axis via the shock at the 
wind/wind or wind/medium interface over comparable length scales with the focusing length
slightly smaller for the case of the single-wind simulation. This small reduction 
is likely due to the fact that mixing of the disk-wind with the post-shock stellar-wind 
will impart additional $z$-directed momentum flux to the flow. Otherwise, 
the resulting shock structures appear very similar in the hydrodynamic cases. 

The differences between the nested-wind and single-wind cases for the MHD 
cases on the other hand are quite striking. We note first that in the absence of
the disk wind, a large rarefied region (a cocoon) opens up along the axis of the wind while
a relatively short MHD nose cone evolves at the working surface of the wind shock with
several vortices appearing along the bow shock. When both winds operate however, 
there is considerable mixing between the two flows giving rise to the filimentary
structures appearing in the region near the axis in place of rarefaction.
In addition, The bow shock has developed a ``shoulder,''
below which an extended and somewhat flattened nose cone protrudes with what appears 
to be a refocusing event similar to those seen in the hydrodynamic cases occuring 
in its interior.  The shoulder begins to develop early in the simulation soon
after the formation of the nose cone and is built up smoothly over the course of
the simulation as separate shedding events occurring near the head of the nose
cone combine, cool, and expand.

\subsection{Energy maps}

\begin{figure*}[ht]
\includegraphics[angle=0,
                 height=6.0in,
                 keepaspectratio=true,
                 trim= 0 0 0 0
                 clip=true]
        {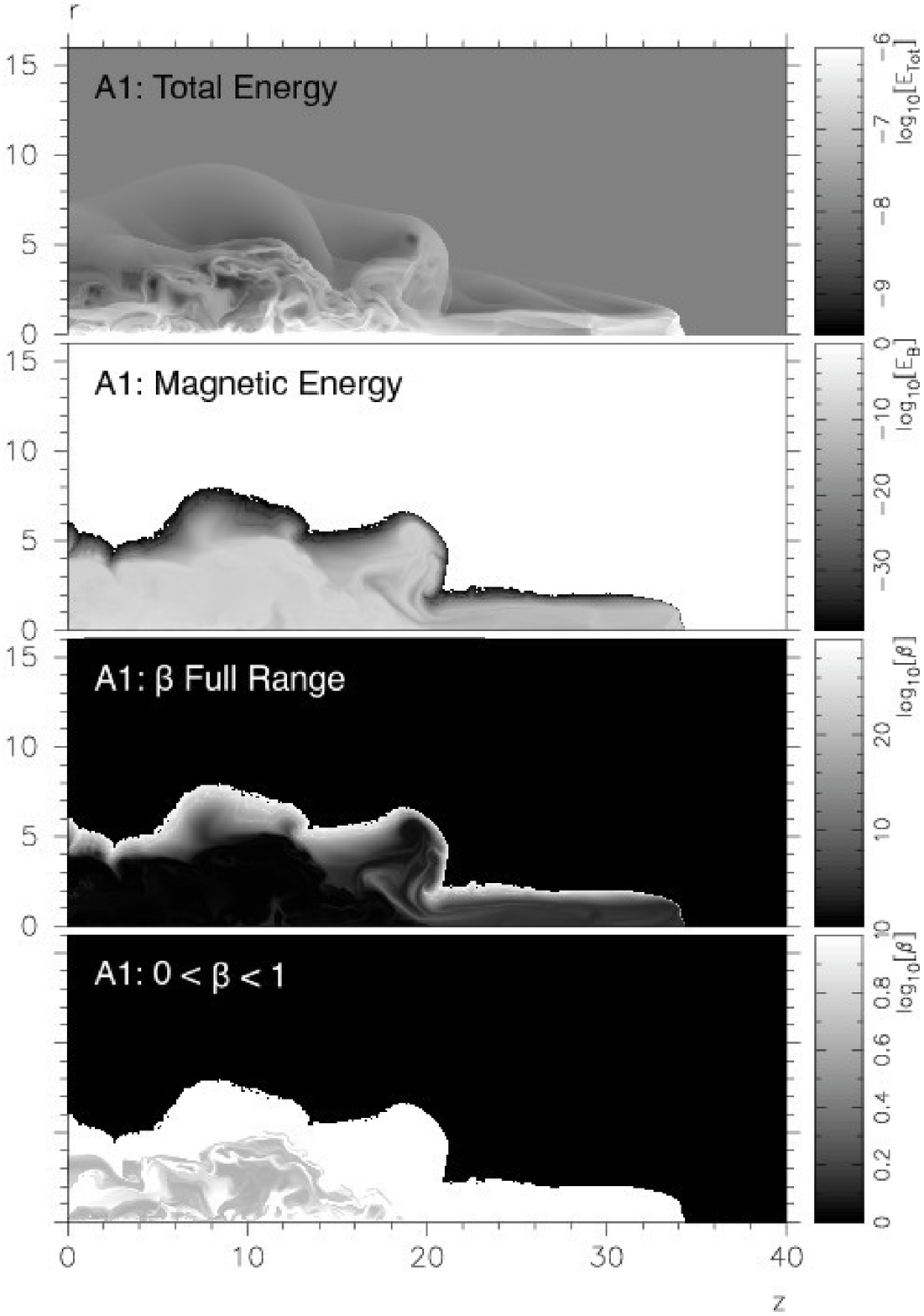}
\caption{ Maps of total energy (top panel), magnetic energy (second panel),  and 
          plasma-$\beta$ (bottom two panels) for the MHD nested wind (run A1).
          For the maps of magnetic energy, the white regions exterior to flow 
          are regions of zero magnetic energy.  For visual clarity, two maps of $\beta$
          are given.  The third panel, shows the full range of variation in 
          $\beta$ while the fourth panel shows a map with color-coding restricted
          to $0<\beta<1$ so that details in the variation of $\beta$ within the 
          beam may be discerned. Pure black regions exterior to the beam have zero 
          magnetic field ($\beta\rightarrow\infty$). In all panels the unit of length
          is $500\au$, and the flow axis is parallel to the image plane.
\label{fig:f3}}
\end{figure*}

\begin{figure*}[ht]
\includegraphics[angle=0,
                 height=6.00in,
                 keepaspectratio=true,
                 trim= 0 0 0 0
                 clip=true]
        {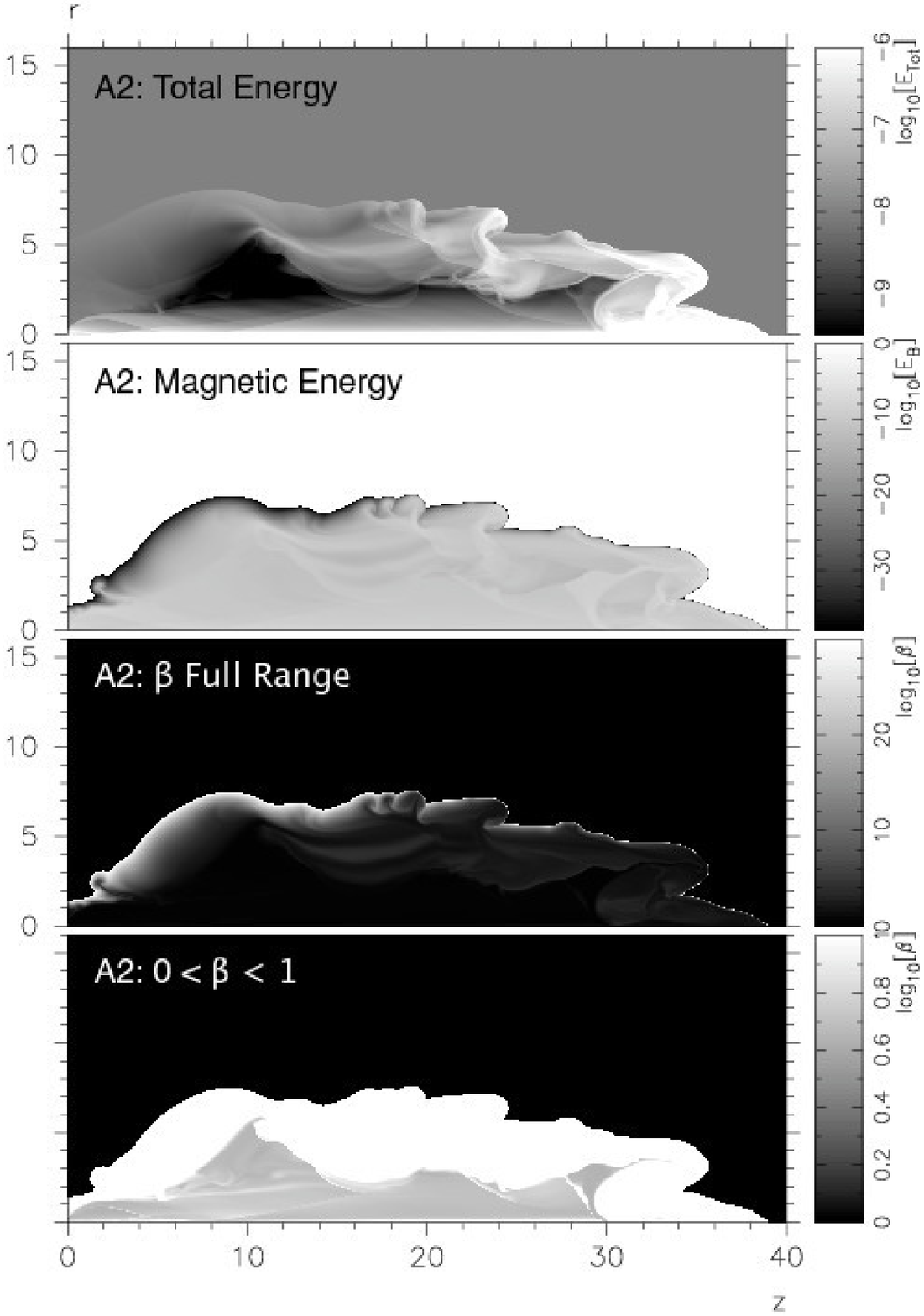}
\caption{ Maps of total energy (top panel), magnetic energy (second panel),  and 
          plasma-$\beta$ (bottom two panels) for the MHD single wind (run A2).
          For the maps of magnetic energy, the white regions exterior to flow 
          are regions of zero magnetic energy.  For visual clarity, two maps of $\beta$
          are given. The third panel, shows the full range of variation in 
          $\beta$ while the fourth panel shows a map with color-coding restricted
          to $0<\beta<1$ so that details in the variation of $\beta$ within the 
          beam may be discerned. Pure black regions exterior to the beam have zero 
          magnetic field ($\beta\rightarrow\infty$). in all panels the unit of length
          is $500\au$ and the flow axis is parallel to the image plane.
\label{fig:f4}}
\end{figure*}

We next compare maps of the total energy density to the 
magnetic energy density, and to maps of plasma-$\beta$ for both the nested-wind and 
single-wind cases. Results for the nested-wind case are shown in figure \ref{fig:f3} 
while the single-wind case is presented in figure \ref{fig:f4}.
%
%
The top two panels of figure \ref{fig:f3} compare total energy density and magnetic 
energy density for the nested-wind case.  We first notice that the map of total 
energy density is quite similar in appearance to the map of density. This is to be 
expected since most of the energy in the system is either thermal or kinetic and thus closely 
traces the density. The map of magnetic energy density has a different appearance.  
Here we note that in the region bounded on the right by the bow shock ``shoulder,'' 
the magnetic energy density is distributed in two roughly uniform layers---distinguished 
by their typical values---throughout the area enclosed and falls off steeply only near 
the wind/medium interface. The energy densities characteristic of the inner layer are 
evidently of order $E_{\mathrm B}\gtrsim 10^{-6}\ergpcc$ while those of the outer layer are 
characterized (very roughly) by energy densities in the range
$0\lesssim E_{\mathrm B}\lesssim 10^{-10}\ergpcc$. It is interesting to note that 
in the extended conical region (an MHD nose cone) to the right of the shoulder, the typical magnetic 
energy density is of order the highest values found in the {\em outer} of the two 
layers to the left of the shoulder. This result is confirmed in the maps of $\beta$ 
given in the bottom two panels of figure \ref{fig:f3} which indicate values of 
$\beta \lesssim 1.0$ throughout much of the inner layer while in both the outer layer 
to the left of the shoulder, and the conical region beyond we find $\beta\gtrsim10$. 
Note that in the maps of $\beta$, the color black is serving double-duty and indicates 
very low values of $\beta$ in the interior of the flow and ``infinite'' values exterior 
to it. For this reason we have presented two maps for each simulation; one in which 
the full range of values of $\beta$ are mapped, and another ``overexposed'' map in 
which values of $\beta$ are restricted to the range $0\le\beta\le1$. In this way we are 
able to show that the very dark interior regions in the panel with unrestricted range 
are regions of low-$\beta$ plasma. Note that the magnetic field in the cocoon/bubble is not in fact very strong and does not play an important role in determing the morphology there. This can be seen by examining the plots of
$\beta$.  In the body of the jet downsteam of the interaction region we see that $\beta$ can be of order 1 and in those regions the field is contributing more significantly to the dynamics. 

The noticeable decrease in magnetic energy density in the extended conical region
is particularly interesting in light of the observation made earlier of what appears 
to be the occurance of a hydrodynamic refocusing event in the interior of the extended 
conical region. These results suggest that while the inertia and over-pressuring of the 
winds initially lead to an expansion into the ambient medium of the winds, 
hoop stresses---coupled with the gradual accumulation of shed material---eventually 
create a bottleneck for the more highly magnetized material in the interior, leading to 
the ejection of relatively less magnetized material, which then exhibits 
refocusing similar to the hydrodynamic wind within this extended region beyond the shoulder. 

The single-wind case shown in figure \ref{fig:f4} tells a very different story.  
While the map of total energy density shows some layering indicative of larger 
energy densities in regions with more material, the magnetic energy is distributed 
uniformly throughout the region of the flow including in the nose cone at the tip. This 
is again confirmed in the bottom two panels of figure \ref{fig:f4} showing the maps of 
$\beta$ for this case.

\subsection{Ambient temperature effects}
\label{sec:cooljets}

Most of the simulations run for this paper assumed a warm ($T_a=10^4 \kelvin)$
ambient medium. Since the type of flows studied here are hypothesized to occur
in the non-ionized mediums of post-AGB stars we present in figure \ref{fig:f6}
the density maps for a pair of nested-wind simulations for the cases of 
MHD and HD flow respectively with ambient temperature set to $T=100\kelvin$. 
The results shown are from runs A3 and B3 in
table \ref{tab:t1} respectively. We note that in the case of the MHD nested-wind,
the same bow shock shoulder, filamentary structures, and extended and flattened
nose cone (with some refocusing again evident), are all present in run A3 as well.
Comparison with the top panel of figure \ref{fig:f2} also indicates that in
spite of the now over-pressured jet impinging upon an ambient environment which
has had its pressure lowered by two orders of magnitude, the flow suffers very
little additional expansion into the medium, suggesting that the hoop stresses
play the predominant role in maintaining collimation for the case of MHD nested
winds. The hydrodynamic nested-wind on the other hand, suffers a significant
amount of expansion when the pressure of the ambient environment is reduced. The 
rarified region along the axis quadruples its radius and is now bordered by a
relatively dense ``beam'' of material consisting of thoroughly mixed post-shock
stellar and disk material, with a characteristic width much greater than the original
dimensions of the flow. This layer is in turn bordered by an even denser layer 
of ambient material plowed up by the flow which, in spite of the shock that
forms at its outer edge fails to warm sufficiently to give rise to significant
radiative cooling. The resulting appearance is that of an adiabatic wind 
``piggybacked'' upon a radiative wind. This impression is further suggested
by the presence of a nose cone near the axis, and a Mach disk above and to its
left.

\subsection{Effect of mass-loss and velocity variations}

\begin{figure*}[ht]
\includegraphics[angle=0,
		 width=3.5in,
                 keepaspectratio=true,
                 trim= 0 0 0 0
                 clip=true]
        {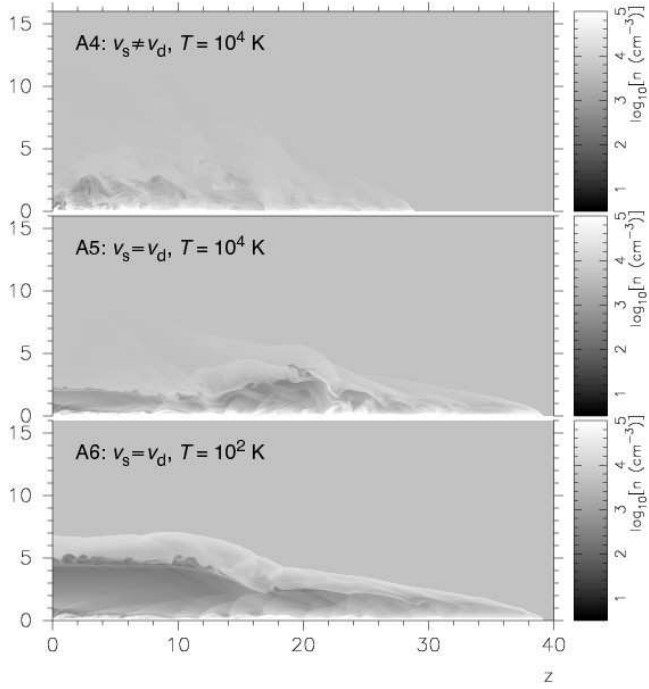}
\caption{Density maps of MHD nested winds for the case of a ``light'' stellar wind. The maps
         shown correspond, from top to bottom, to runs A4, A5, and A6, respectively 
         (See table \ref{tab:t1}). In all panels the unit of length is $500\au$ and the flow 
         axis is parallel to the image plane.
\label{fig:f7}}
\end{figure*}

It is expected for both the post-AGB and YSO systems that both the stellar and 
disk winds are expected to have---to order of magnitude---comparable velocities and 
mass-loss rates. Still, it is important to investigate how differences
in these quantities within the expected range of variation are likely to 
affect the appearance and physics of the flows. With this in mind we have 
produced the three simulations for which we present late-time density maps
in figure \ref{fig:f7}. In the top panel of this figure we show the results
of run $A4$ in which the velocities of the flows---$150 \kmpersec$ for the
stellar wind and $50 \kmpersec$ for the disk wind---are unchanged from our 
standard run A1, but the mass-loss rate for the stellar wind has been
reduced by one order of magnitude relative to the mass-loss rate for the disk
wind. Because the temperature of the stellar wind is determined by its velocity
and the two parameters $\sigma$ and $\beta_m$, all of which are unchanged from
run A1, the effect is to lower the mean density and therefore the pressure
of the stellar wind. This allows material from the slower disk wind to expand 
into, mix with, and disrupt the flow of the stellar wind material, creating
a turbulent, filamentary field in the region near the axis which, in run A1
is relatively evacuated.  

In the second panel of figure \ref{fig:f7} we show results from run A5 in 
which we have now set the velocities to equal values ($100 \kmpersec$). We 
see that because of the reduced pressure of the stellar wind, the disk wind is 
still able expand into the region of the stellar wind, and mix with the material 
there, but because the velocities are equal, this expansion is less disruptive 
to the flow there.  We see a relatively more uniform density field in this region. 
From comparing these panels to the first of figure \ref{fig:f2},
in which we also noted the presence of filamentary density structures, we may
conclude that differences in both mass-loss rates and flow velocities 
contribute to determining the degree to which these filamentary density 
structures arise, and to the characteristic length scales associated with them,
in the region of the flow near the axis. In the bottom panel of figure \ref{fig:f7}, we 
present the result of run A6 which is identical to run A5 with the exception that 
the temperature of the ambient medium has been reduced from $10^4 \kelvin$ to 
$100\kelvin$.  It is evident in this panel that the effect of this reduction, 
is to allow for greater expansion of the entire over-pressured flow into the 
surrounding medium with otherwise little qualitative alteration in the appearance 
of the flow.

\subsection{Mapping the emission}

\begin{figure*}[ht]
\includegraphics[angle=0,
		 width=6.0in,
                 keepaspectratio=true,
                 trim= 0 0 0 0
                 clip=true]
        {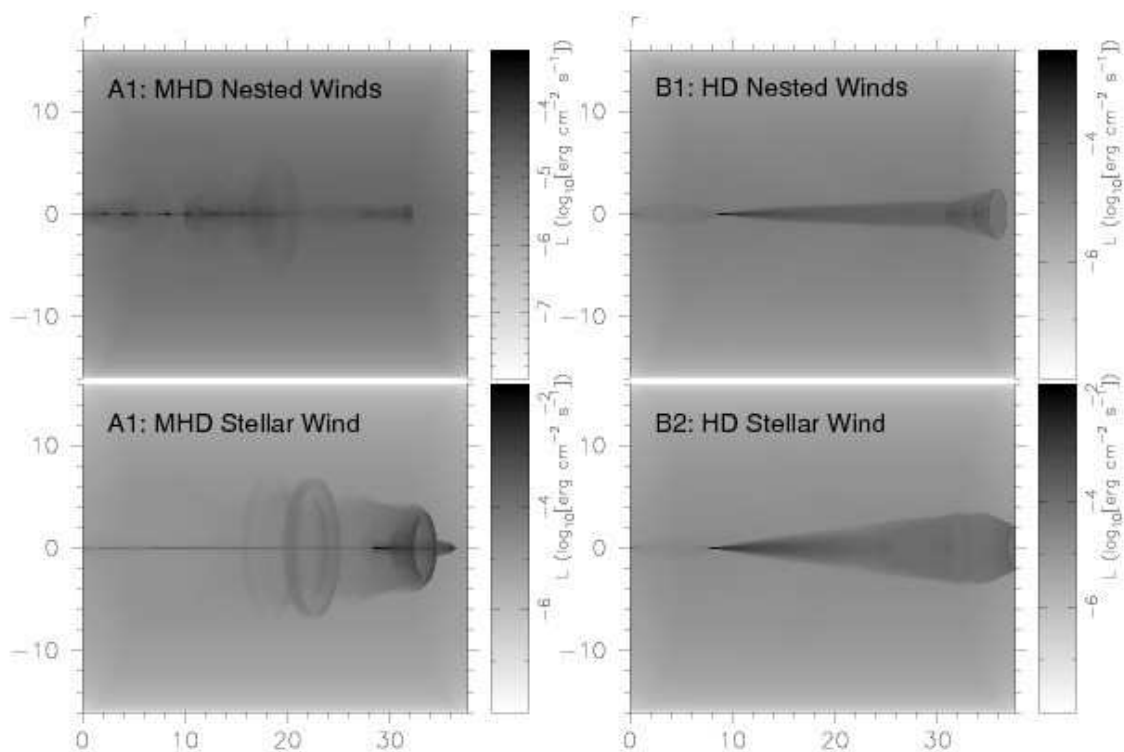}
\caption{Synthetic maps of radiative emission  for nested and single wind simulations. 
         The panels in the left column of the figure are from the MHD nested wind (top) 
         and stellar winds (bottom) respectively (runs A1 and A2).  The panels in the 
         right column are from the HD nested (top) and stellar (bottom) winds respectively 
         (runs B1 and B2). In all panels the unit of length is $500\au$ and the flow axis 
         makes a $20^\circ$ angle with respect to the image plane.
\label{fig:f8}}
\end{figure*}

In an attempt to approximate roughly how our interacting winds might 
appear on the sky, we present in figure \ref{fig:f8} synthetic maps of emission
for both our nested-wind and single wind cases as obtained in the MHD and HD
runs. 
 The intensity shown, which does not distinguish among cooling lines,
 was determined according to:
 \begin{equation}
 I_{i,j,k} = \Sigma_k n_{i,j,k}^2\Lambda(T_{i,j,k}),
 \end{equation}
 where $i$,$j$, and $k$ refer to the $x$,$y$ and $z$ directions in the final data cube 
 created by rotating $n(r,z)$ and $T(r,z)$ about the axis of symmetry, and $\Lambda$ is 
 the cooling function. A projection angle of $20^\circ$ is assumed in all panels.
The left column shows the results for the MHD runs with the nested-winds
shown in the top panel and the single-wind shown below. The column on the right
shows the corresponding results for the HD case. The maps were generated by revolving
the 2.5D data about the axis of symmetry, ``tilting'' the symmetry axis by
a projection angle of $20^\circ$, and subsequently summing along lines of sight 
and projecting the results on to the image plane. The results are for the total 
radiative emission. No attempt was made to distinguish among lines of emission. 
The images show a marked distinction between the MHD and HD cases.  While the 
HD simulations give rise to smooth conical segments of emission which broaden with 
distance, the MHD results instead give rise to emission divided by primarily on-axis 
features and ring-like structures centered on the symmetry axis. We also note 
an interesting difference that appears in the on-axis emission of the nested-winds 
as contrasted with the stellar-wind case. The emission features on both axes are 
both quite narrow, but while the on-axis emission for the stellar wind is quite 
smooth, the corresponding MHD emission shows small but distinct knots of emission 
distributed along the left half of the axis and vanishing approximately where we 
earlier noted a bottleneck in magnetic energy arose. While it is possible
that the smooth emission seen in both panels could be in part an 
artifact of the numerical method employed for handling the cylindrical symmetry 
of the problem, the appearance of the knots in the nested-wind case and their 
absence in the stellar-wind case suggests that these features are ``real.'' Given 
that well-aligned knots of emission are routinely observed in optical HH jets, 
these results, while preliminary, provide cause for asking if these knots of 
emission are a consequence of interacting nested winds in the environments of YSO's.

\begin{figure*}[ht]
\includegraphics[angle=0,
		 width=6.0in,
                 keepaspectratio=true,
                 trim= 0 0 0 0
                 clip=true]
        {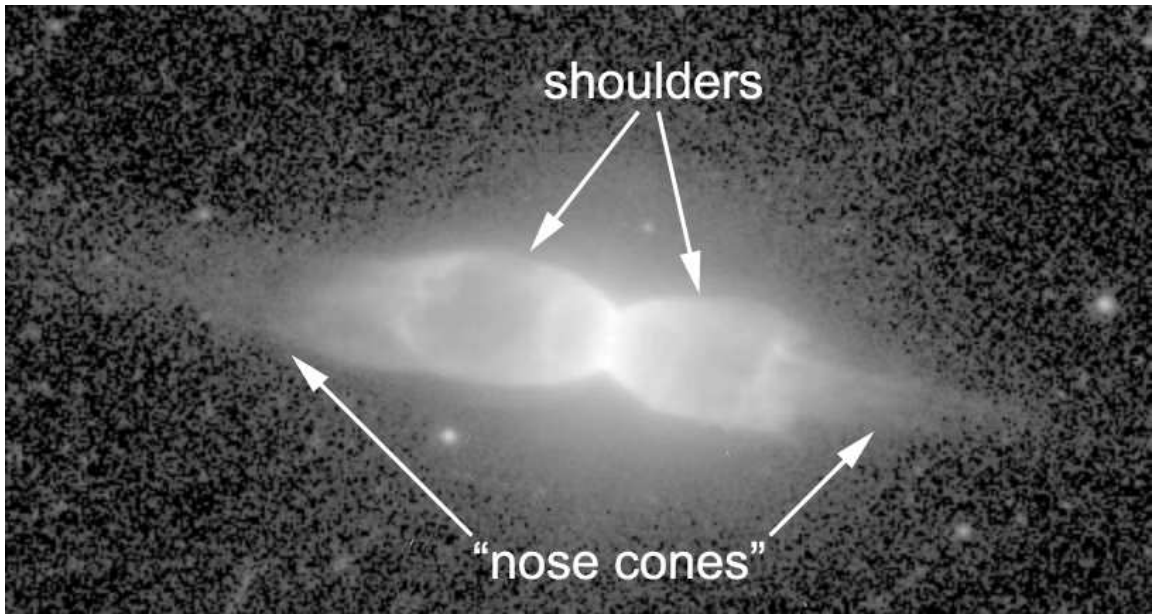}
\caption{An image of Hen 2-320 taken with the HST/WFPC2/PC camera through the F659N filter 
         and displayed logarithmically (credit: NASA and HST proposal GO8345, R. Sahai, P.I.). 
         See section 3 for a discussion of the identified features.'' 
\label{fig:f9}}
\end{figure*}

The feature that most distinguishes the MHD nested-wind from the other scenarios 
considered here is the development of the bow shock shoulder mentioned above. 
This feature was plainly evident in both the warm and cool nested-wind MHD simulations 
presented here, and would serve as a marker for determining whether an observed object 
has been formed from simultaneously operating magnetized disk and stellar winds. Before 
we can robustly employ such a feature as a means of identifying objects as candidates for 
formation by nested-winds, it will be necessary to study the formation of the shoulder 
in greater detail. For example, it will be necessary to determine how the formation and 
appearance of the shoulder varies with such parameters as the disk opening angle, ratios 
of disk and stellar flow velocities, and ratios of mass-loss rates.  (In particular, it 
is evident from figure \ref{fig:f7}, that the shoulder is much more subtle in appearance 
when the mass-loss rate of the star is significantly lower than that of the disk.) But 
to illustrate what we have in mind, we present as a case in point the Hubble Space 
Telescope (HST) image of the planetary nebula Hen 2-320 in figure \ref{fig:f9}. Though 
other interpretations are possible, the right lobe of this object exhibits features 
that are qualitatively similar to our MHD nested-wind simulations. We see---indicated
by the arrows labeled ``shoulders''---an expanded region of the flow nearer to the nebular core, 
and---indicated by the arrows labeled ``nose cones''---a narrower and conical extended region 
to its right. We also note a brightening of emission at the location where the expanded 
region gives way to the narrow conical region.  This brightening is reminiscent of the 
large ring of (synthetic) emission seen in the upper-left panel of figure \ref{fig:f8} 
which we note coincides with the location of the bow shock shoulder that arises in this 
simulation.

\section{Discussion and Conclusion}
\label{sec:conc}
We have presented a series of magnetohydrodynamic (MHD) and hydrodynamic (HD) 
simulations of the flow arising from the imposition of a simultaneous pairing of 
nested, steady-state, diverging, interacting, radiative winds on a homogeneous
quiescent circumstellar environment. The study was motivated observationally
by studies indicating the simultaneous presence of both disk and stellar winds
in the environments of proto-planetary nebulae (PPN's), planetary nebulae (PN's) and
young stellar objects (YSO's) (Livio 1997 and references therein); and theoretically
by the consensus view that disk winds are magnetically launched, and that 
magnetocentrifugal launch mechanisms acting impulsively over time scales
corresponding to the rapid evolution of a poloidally magnetized object are favored
to explain the origin of gamma-ray bursts (Piran 2005) and thus---by extension---may 
plausibly be conjectured to operate in other analogous contexts as well,
such as in the environments of the central post-AGB stars associated with 
PN's and PPN's.

The results of our simulations demonstrate that the physical processes predominantly 
responsible for maintaining collimation in the outflows differ between the 
MHD and HD cases, with magnetic hoop stresses being the chief agent of collimation
for the MHD runs, while axially-directed refocusing by the shock arising at the 
wind/environment interface serves a similar purpose in the case of the HD outflows.
We also find that the density structures exhibited at late times in the
simulations differ significantly both when comparing MHD outflows to HD outflows
and when comparing either MHD or HD nested-wind outflows to outflows in which
the disk wind is absent. The differences between the nested and single, stellar-wind
MHD outflows are the most striking, indicating that the presence of the
disk wind has a profound influence on the appearance of both the near-axis
regions of the beam and the shape of the bow shock, while the differences 
between the nested and stellar-wind HD outflows are more subtle showing topologically
similar density structure differing mainly in geometric distortions of the bow shock. 
We have also shown that these results are not particularly sensitive to differences
in temperature of the environment sufficient to distinguish between ionized and 
un-ionized gas; and that significant disparities in the mass-loss rates parameterizing
the stellar versus the disk winds lead to expansion of the denser disk wind material
into the region of the stellar wind and the subsequent turbulent disruption of the 
flow there, and that this effect is most evident when the disk and stellar wind speeds
are substantially different. 

The work of Akashi 2007, Akashi \& Soker 2008 and Akashi et al 2008 bears noting in relation to the present study.  In those studies hydrodynamic simulations of a wide jets ($\theta=30^o$) burrowing into an AGB wind were presented including radiative cooling below 9000K.  While magnetic fields were not included in those simulations the results demonstrate the rich variety of features that can be produced in these systems.  The results of Akashi and the current study are generally in accord. For example on the issue of temperature in the ambient medium we find that only the MHD runs were relatively unaffected by a change in ambient temperature as one might expect because hoop stresses dominate lateral expansion.  The hydrodynamic runs, which are relevant to the Akashi simulations do show significant differences between the cool and hot ambient media in ways that are qualitatively similar to what Akashi saw.  It is also noteworthy that Akaski \& Soker have been able to create features with protrusions at the head which create the front lobes seen in some PN's without magnetic fields.  In MHD the nose cone is a conical region formed when the jet shock stands off at some distance from the bow shock and jet material flowing into the shock does not, effectively, escape into the cocoon.  This occurs in MHD jets due to hoop stresses from the toroidal field which restrict the shocked jet gas from radial motion.  Akashi \& Soker and the current simulations show that conical features at the head of the jet can occur in pure hydrodynamic simulations (though we note it will be a source of confusion to call these nose-cones).  We note that further work needs to be done in 3-D in both hydro and MHD to explore the stability of these conical features.  Finally we note that the simulations of Akashi \& Soker developed dense equatorial rings such as have been observed in some PN and PPN (Hen 2-320). The ability for wide jets to create such features is an attractive feature of the models.  In the current simulations we did not observe such dense rings however this is likely a deficiency which occurs due to the size of our inflow region and our choice of a constant density ambient medium. We note that a dense ring was observed to form in the jet/AGB wind models of Garcia-Arrendendo \& Frank 2006 where the jet launching region was resolved.

Finally, synthetic maps of emission suggest that 
fundamentally differing morphologies should be expected between outflows
arising from purely HD winds and those arising from MHD winds irrespective of whether
both the stellar and disk winds, or just the stellar winds are operating in the
environment. They also indicate that when both disk and stellar winds are operating,
knots of emission reminiscent of those seen in the highly collimated optical jets
associated with YSO's appear on the axis of symmetry. That these knots are 
conspicuously absent when only the MHD stellar wind is operating suggests that they
may not be an artifact of the axial symmetry imposed by the simulation. 
This result in particular, points out the need for further work involving fully 
3D simulations. 

\acknowledgements
The authors thank Alexei Poludnenko, Joel Kastner, Patrick Hartigan, 
Pat Huggins, and Raghvendra Sahai for their insights.

This material is based upon work supported by NASA under Award No. NM0710076, through
the Jet Propulsion Laboratory Spitzer Space Telescope Theory grant 051080-001; by the
National Science Foundation under Award No.'s AST-0507519, and Hubble Space Telescope 
theory grant 11251; and by the Department of Energy under Award No. DE-FC52-92SF19460.
OD acknowledges NSF grant AST-0607111.

\clearpage

%
%
%
%
%
%
%
%
\end{document}